\documentclass[twocolumn,showpacs,aps,prd,floatfix]{revtex4}
\usepackage{graphicx}
\usepackage{dcolumn}
\usepackage{multirow}
\usepackage{amsmath}
\usepackage{epsfig}
\usepackage{color}
\usepackage{verbatim} 
\usepackage{rotating} 

\input{pubboard/babarsym}

\newcommand{\commenthww}[1]{}

\def\BBbar{\BB}
\def\qq{\qqbar}

\def\er #1 #2{ $#1 \pm #2$ \xspace}
\def\om{\ensuremath{\omega}\xspace}

\def\Emiss{\ensuremath{E_{\rm miss}}\xspace}
\def\pmiss{\ensuremath{p_{\rm miss}}\xspace}
\def\vpmiss{\ensuremath{\vec{p}_{\rm miss}}\xspace}
\def\Mmiss2Emiss{\ensuremath{m_{\rm miss}^2/(2 E_{\rm miss})}\xspace}
\def\cosBY{\ensuremath{\cos\theta_{BY}}\xspace}
\def\cosThrust{\ensuremath{\cos\Delta\theta_{\rm thrust}}\xspace}
\def\L2{\ensuremath{L_2}\xspace}
\def\R2{\ensuremath{R_2}\xspace}
\def\Yvtx{\ensuremath{Y_{\rm vtx}}\xspace}

\def\ctl{\ensuremath{\cos\theta_{W\ell}}\xspace}

\def\m3pi{\ensuremath{m_{3\pi}}\xspace}
\def\mES{\ensuremath{m_{\rm ES}}\xspace}
\def\DeltaE{\ensuremath{\Delta E}\xspace}
\def\q2{\ensuremath{q^2}}
\def\dBdq2{\ensuremath{\Delta{\cal B}/\Delta q^2}\xspace}

\def\gevSq{\ensuremath{\rm \,Ge\kern -0.08em V^2}}
\def\rad{\ensuremath{\rm \, rad}} 

\def\Bpilnu{\ensuremath{B \rightarrow \pi\ell\nu}\xspace}
\def\Brholnu{\ensuremath{B \rightarrow \rho\ell\nu}\xspace}

\def\Bptoomegalnu{\ensuremath{B^{+} \rightarrow \omega\ell^+\nu}\xspace}
\def\Btoomegalnu{\Bptoomegalnu}
\def\BtoXulnu{\ensuremath{B \rightarrow X_u\ell\nu}\xspace}
\def\BtoXclnu{\ensuremath{B \rightarrow X_c\ell\nu}\xspace}

\def\BtoDstarlnu{\ensuremath{B \rightarrow \Dstar\ell\nu} \xspace}
\def\BtoDlnu{\ensuremath{B \rightarrow D\ell\nu} \xspace}

\def\qq{\ensuremath{q \bar{q}}\xspace}

\def\btodstarcharges{\ensuremath{B^0 \ra D^{*-}\ell^{+}\nu}\xspace}

\def\bclnu{\ensuremath{b \to c\ell\nu}\xspace}

\def\mpl #1 #2 #3{Mod.~Phys.~Lett.~{\bf#1},\ #2 (#3)\xspace}
\def\npb  #1 #2 #3{Nucl.~Phys.~B~{\bf#1},\ #2 (#3)\xspace}
\def\plb  #1 #2 #3{Phys.~Lett.~B~{\bf#1},\ #2 (#3)\xspace}
\def\pr   #1 #2 #3{Phys.~Rep.~{\bf#1},\ #2 (#3)\xspace}
\def\prd  #1 #2 #3{Phys.~Rev.~D~{\bf#1},\ #2 (#3)\xspace}
\def\prl  #1 #2 #3{Phys.~Rev.~Lett.~{\bf#1},\ #2 (#3)\xspace}
\def\RMP  #1 #2 #3{Rev.~Mod.~Phys.~{\bf#1},\ #2 (#3)\xspace}
\def\zpc  #1 #2 #3{Z.~Phys.~C~{\bf#1},\ #2 (#3)\xspace}
\def\nim  #1 #2 #3{Nucl.~Instrum.~Methods~{\bf#1},\ #2 (#3)\xspace}
\def\nima  #1 #2 #3{Nucl.~Instrum.~Methods~A.{\bf#1},\ #2 (#3)\xspace}
\def\epjc #1 #2 #3{Euro.~Phys.~Jour~{\bf#1},\ #2 (#3)\xspace}
\def\rmp #1 #2 #3{Rev.~Mod.~Phys~{\bf#1},\ #2 (#3)\xspace}
\def\npbps #1 #2 #3{Nucl.~Phys.~B.~proc.~suppl~{\bf#1},\ #2 (#3)\xspace}
\def\progtp #1 #2 #3{Prog.~Theo.~Phys~{\bf#1},\ #2 (#3)\xspace}

\long\def\inst#1{\par\nobreak\kern 4pt\nobreak
    {\em #1}\par\vskip 10pt plus 3pt minus 3pt}

\def\qq{\ensuremath{q\bar{q}}\xspace}

\def\Bpilnu{\ensuremath{B \rightarrow \pi \ell\nu}\xspace}
\def\Brholnu{\ensuremath{B \rightarrow \rho \ell\nu}\xspace}

\def\bclnu{\ensuremath{B \rightarrow X_c\ell\nu}\xspace}

\def\Emiss{\ensuremath{E_{\rm miss}}\xspace}
\def\pmiss{\ensuremath{p_{\rm miss}}\xspace}

\def\mES{\ensuremath{m_{\rm ES}}\xspace}
\def\DeltaE{\ensuremath{\Delta E}\xspace}
\def\cosBY{\ensuremath{\cos\theta_{\rm BY}}\xspace}

\usepackage{subfigure}

\newcommand{\BABARPubYear}    {12}
\newcommand{\BABARPubNumber} {005}
\newcommand{\SLACPubNumber}{15029}

\begin{document}

  \preprint{\babar-PUB-\BABARPubYear/\BABARPubNumber} 
  \preprint{SLAC-PUB-\SLACPubNumber} 
  
  \begin{flushleft}
    \babar-PUB-\BABARPubYear/\BABARPubNumber\\
    SLAC-PUB-\SLACPubNumber\\
  \end{flushleft}
  
\title{\large\bf Branching fraction measurement of \Btoomegalnu decays}
\author{J.~P.~Lees}
\author{V.~Poireau}
\author{V.~Tisserand}
\affiliation{Laboratoire d'Annecy-le-Vieux de Physique des Particules (LAPP), Universit\'e de Savoie, CNRS/IN2P3,  F-74941 Annecy-Le-Vieux, France}
\author{J.~Garra~Tico}
\author{E.~Grauges}
\affiliation{Universitat de Barcelona, Facultat de Fisica, Departament ECM, E-08028 Barcelona, Spain }
\author{A.~Palano$^{ab}$ }
\affiliation{INFN Sezione di Bari$^{a}$; Dipartimento di Fisica, Universit\`a di Bari$^{b}$, I-70126 Bari, Italy }
\author{G.~Eigen}
\author{B.~Stugu}
\affiliation{University of Bergen, Institute of Physics, N-5007 Bergen, Norway }
\author{D.~N.~Brown}
\author{L.~T.~Kerth}
\author{Yu.~G.~Kolomensky}
\author{G.~Lynch}
\affiliation{Lawrence Berkeley National Laboratory and University of California, Berkeley, California 94720, USA }
% ...
\collaboration{The \babar\ Collaboration}
\noaffiliation

\begin{abstract}

We present a measurement of the \Btoomegalnu branching fraction 
based on a sample of 467 million \BB pairs recorded by the \babar\
detector at the SLAC \pep2 \epem collider.  
We observe $1125 \pm 131$ signal decays, corresponding to a
branching fraction of
$\BR(\Btoomegalnu) = (1.21 \pm 0.14 \pm 0.08) \times 10^{-4}$, 
where the first error is statistical and the second is systematic.
The dependence of the decay rate on \q2, 
the invariant mass squared of the leptons, 
is compared to QCD predictions of the 
form factors based on a quark model and light-cone sum rules.

\end{abstract}


\maketitle  

\section{Introduction}
Most theoretical and experimental studies of exclusive \BtoXulnu decays have focused on \Btopilnu decays, while
\Brholnu and \Btoomegalnu \cite{chgConjugate} decays involving 
the vector mesons $\rho$ and \om
have received less attention.  
Here $\ell$ is an electron or muon, and $X$ refers to a hadronic state, with the subscript $c$ or $u$ signifying
whether the state carries charm or is charmless.
Measurements of the branching fraction of \Btorholnu are impacted by an  
irreducible \BtoXulnu background, typically the dominant source of systematic 
uncertainty. 
In studies of \Btoomegalnu that background can be suppressed to a larger 
degree, since the \om width is about 15 times smaller than that of the $\rho$.
Extractions of the CKM matrix element \Vub\ from \Btoomegalnu and \Brholnu decay rates
have greater uncertainties than those from
\Bpilnu, due to higher backgrounds and more complex form-factor dependencies.
The persistent discrepancy between  \Vub measurements based on inclusive and exclusive 
charmless decays is a motivation for 
the study of different exclusive \BtoXulnu decays \cite{KowalewskiPDG, HFAG2010}. 

Measurements of \BR(\Btoomegalnu) have been reported by Belle 
\cite{Schwanda:2004fa,BelleOmegalnu2008}; a measurement by \babar\ has been performed on 
a partial dataset \cite{Anders}.  
In this analysis we use the full \babar\ dataset to measure the total branching fraction \BR(\Btoomegalnu) 
and partial branching fractions $\Delta\BR(\Btoomegalnu)/\Delta\q2$ in five \q2\ intervals,
where \q2\ refers to the momentum transfer squared to the lepton system.

The differential decay rate for \Bptoomegalnu is given by \cite{burchat}
\begin{eqnarray}
\frac{{\rm d}\Gamma(\Bptoomegalnu)}{{\rm d}\q2}
= |V_{ub}|^2 \,  \frac{G^2_F \,q^2 \, p_\om}{96\pi^3 m_B^2 c_V^2}  \nonumber \\
\times{} \bigg[ |H_0|^2 + |H_+|^2 + |H_-|^2 \bigg] \ , 
\label{eqn:decrateq2}
\end{eqnarray}
where $p_{\om}$ is the magnitude of the \om momentum in the $B$ rest frame,
$m_B$ is the $B$ mass, and 
$G_F$ is the Fermi coupling constant.  
The isospin factor $c_V$ is equal to $\sqrt{2}$ for \Btoomegalnu \cite{Ball05}.  
As described in a related \babar\ paper \cite{VubJochen}, the three helicity functions $H_0$, $H_+$, and $H_-$ can be expressed in terms of two
axial vector form factors $A_1$ and $A_2$ and one vector form factor $V$,
which describe strong interaction effects,   
\begin{align*}
H_{\pm}(q^2) = (m_B+m_{\om}) \bigg[ A_1(q^2)\mp \frac
  {2m_B\,p_{\om}}{(m_B+m_{\om})^2} V(q^2)\bigg]  \ , \nonumber  \\
H_{0}(q^2) = \frac{m_B+m_{\om}}{2m_{\om} \sqrt{q^2}}   
\times \bigg[ ( m_B^2 - m_{\om}^2 - q^2) A_1(q^2) \nonumber \\
-  \frac {4 m_B^2 \,p_{\om}^2} {(m_B+m_{\om})^2} A_2(q^2)\bigg]  .  
\end{align*}
\noindent
We compare the measured \q2\ dependence of the decay rate with form factor predictions 
based on light-cone sum rules (LCSR)~\cite{Ball05} and the ISGW2 quark model~\cite{isgw2}.
We also use these form factor calculations and the measured branching fraction to extract \Vub.

\section{Detector, Data Set, and Simulation} 

The data used in this analysis were recorded with the 
\babar\ detector at the \pep2 $e^+e^-$ collider
operating at the \Y4S resonance.  
We use a data  sample of 426 \invfb,  
corresponding to  
($467 \pm 5$)
million produced \BB pairs.  In addition, we use 44
\invfb of data collected 40 \mev below the \BB production
threshold.  This off-resonance sample is used to validate the 
simulation of the non-\BB\ contributions whose principal source is 
\epem\ annihilation to \qqbar\ pairs, 
where $q=u,d,s,c$.  

The \pep2\ collider and \babar\ detector have been described in detail
elsewhere~\cite{NIM}. 
Charged particles are reconstructed in a five-layer silicon tracker
positioned close to 
the beam pipe and a forty-layer drift chamber.  Particles of different
masses are distinguished by their ionization energy loss in the
tracking devices and by a ring-imaging Cerenkov 
detector.
Electromagnetic showers from electrons and photons are measured in a
finely segmented CsI(Tl) calorimeter.   
These detector components are embedded in a $1.5$ $\mathrm{T}$
magnetic field of a superconducting solenoid; its steel flux return
is segmented and instrumented with planar resistive plate chambers and
limited streamer tubes to detect muons that penetrate the magnet
coil and steel.  

We use Monte Carlo (MC) techniques~\cite{Lange:EvtGen,jetset} to simulate the production
and decay of \BB\ and \qqbar\ pairs and the detector response~\cite{geant4}, 
to estimate signal and background efficiencies and resolutions,
and to extract the expected signal and background distributions. 
The size of the simulated sample of generic \BB\ events exceeds the \BB\ data sample by
about a factor of three, while the MC samples for inclusive and exclusive $B \to X_u \ell \nu$ decays 
exceed the data samples by factors of 15 or more.  The MC sample for
\qqbar\ events is about twice the size of the \qqbar\ contribution in the \FourS\ data. 

The MC simulation of semileptonic decays uses the same models as in a recent 
\babar\ analysis \cite{VubJochen}.  
The simulation of inclusive charmless semileptonic decays \BtoXulnu is based on predictions of a heavy quark expansion \cite{DeFazioNeubert} for the differential decay rates. 
For the simulation of \Btopilnu decays we use the ansatz of \cite{Becirevic} for the \q2 dependence, 
with the single parameter $\alpha_{BK}$ set to the value determined in a previous
\babar\ analysis \cite{cotePilnu}.  
All other exclusive charmless semileptonic decays \BtoXulnu, including the signal, 
are generated with form factors determined by LCSR \cite{Ball05, Ball05scalar}.  
For \BtoDlnu and \BtoDstarlnu decays we use parameterizations of the form factors \cite{Isgur1989, CLN} based on heavy quark effective theory; 
for the generation of the decays $B \to D^{**}\ell\nu$, 
we use the ISGW2 model \cite{isgw2}.

\section{Candidate Selection}

In the following, we describe the selection and kinematic reconstruction of signal candidates, the definition of the various background classes, and the application of neural networks to further suppress these backgrounds.

The primary challenge in studying charmless semileptonic 
$B$ decays is to separate signal decays from Cabibbo-favored \BtoXclnu decays, which
have a branching fraction approximately 50 times larger than that of \BtoXulnu.  
A significant background also arises due to multi-hadron continuum events.  

Based on the origin of the candidate lepton we distinguish three categories 
of events: 1) {\it Signal} candidates with a charged lepton from a true \Btoomegalnu decay; 
2) {\it \BB\ background} with a charged lepton from all non-signal \BB events; 
3) {\it Continuum background} from $\epem \to \qq$ events.  
The \om meson is reconstructed in its dominant decay, $\om \rightarrow \pi^+\pi^-\pi^0$. 
For each of the three categories of events we distinguish correctly reconstructed 
$\om \rightarrow \pi^+\pi^-\pi^0$ decays 
(true-\om ) from combinatorial-\om candidates, for which at least one of the 
reconstructed pions originates from a particle other than the \om.

\subsection{Preselection}

Signal candidates are selected from events with at least four charged tracks, since a  
\Btoomegalnu decay leaves three tracks and the second $B$ in the event is expected to produce at 
least one track.  
The magnitude of the sum of the charges of all reconstructed tracks is required to be less than two, helping
to reject events with at least two undetected particles.

The preselection places requirements on the reconstructed lepton, \om meson, and neutrino 
from the \Btoomegalnu decay.  
At this stage in the analysis we allow for more than one candidate per event.

The lepton is identified as either an electron or muon.  
The electron identification efficiency is greater than 90\% and constant as a function of momentum 
above 1 \gev, while the muon identification efficiency is between 65\%--75\% for momenta of 1.5--3 \gev.  
The pion misidentification rates are about 0.1\% for the electron selector and
1\% for the muon selector.
The lepton is required to have 
a momentum in the center-of-mass (c.m.) frame greater than 1.6 \gev.  
This requirement significantly reduces the background from hadrons
that are misidentified as leptons, and also removes a large fraction
of true leptons from secondary decays or photon conversions and from
\bclnu\ decays. 
The acceptance of the detector for leptons covers momentum polar angles in the range 
$0.41 \leq \theta \leq 2.54 \rad$.

For the reconstruction of the decay $\om \rightarrow \pi^+\pi^-\pi^0$,   
we require that the candidate charged pions are not identified as leptons or kaons.
The reconstructed \om  mass must be in the range
$680 < \m3pi < 860 \mev$, and the  
\piz candidate is required to have an invariant mass of 
$115 < m_{\gamma\gamma} < 150 \mev$.   
To reduce combinatorial \om background, we require minimum
momenta for the three pion candidates,  
  $p_{\pi^\pm}>200$ \mev and  $p_{\pi^0}  >400$ \mev, and also 
 energies of at least 80 \mev for  photons from
  the $\pi^0 $ candidate.

The charged lepton candidate is combined with an \om candidate to
form a  so-called $Y$ candidate. 
The charged tracks associated with the $Y$ candidate are fitted to a
common vertex \Yvtx. This vertex fit must yield a $\chi^2$~probability 
Prob$(\chi^2, \Yvtx) >0.1$.  
To further reduce backgrounds without significant signal losses, we
impose two-dimensional restrictions on the momenta of the lepton
and \om.  
Each $Y$ candidate must satisfy  at least one of the following 
conditions on the c.m. momentum of the lepton and \om:
  $p^*_\om>1.3 \gev$, or $p^*_\ell>2.0 \gev$, or 
  $p^*_\ell+p^*_\om > 2.65 \gev$, 
where quantities with an asterisk refer to the c.m.\ frame.    
These requirements reject background candidates that are inconsistent with the phase space of the signal decay.
The condition $|\cosBY|\leq 1.0$, where 
$\cosBY=(2 E^*_B E^*_{Y} - M_B^2 -M_{Y}^2)/(2 p^*_B\ p^*_{Y})$ is 
the cosine of the angle between the momentum vectors of the $B$ meson and the $Y$ candidate, 
 should be fulfilled 
for a well-reconstructed $Y$ candidate originating from a signal decay~\cite{cleo}.
The energy $E^*_B$ and momentum $p^*_B$  of the $B$ meson are not measured event by event.  Specifically, $E^*_B=\sqrt{s}/2$, where
${\sqrt s}$ is the c.m. energy of the colliding beams, 
and the $B$ momentum is derived as $p^*_B=\sqrt{E^{*2}_B - m^2_B}$. To allow for the finite resolution of the detector, 
we impose the requirement $-1.2 < \cosBY < 1.1$.

The neutrino four-momentum is inferred from  the missing energy and momentum
of the whole event:  
$(\Emiss, \vec{p}_{\rm miss}) = (E_{\epem}, \vec{p}_{\epem}) - (\sum_i{E_i}, \sum_i{\vec{p}_i})$,  
where $E_{\epem}$ and $\vec{p}_{\epem}$ are the energy and momentum of the colliding beam particles, and the sums are performed over 
all tracks and all calorimeter clusters without an associated track. 
If all tracks and clusters in an event are well-measured, and there are no 
undetected particles besides a single neutrino, then 
the measured distribution of the missing mass squared,
$m^2_{\rm miss} = E_{\rm miss}^2 - {p}_{\rm miss}^2$,  peaks at zero. 
We require the reconstructed neutrino mass to be consistent with zero, 
$|m_{\rm miss}^2/(2E_{\rm miss})| < 2.5 \gev$, and the missing momentum 
to exceed 0.5 \gev.
The polar angle of the missing momentum vector is also required to pass through
the fiducial region of the detector, 
$0.3<\theta_{\rm miss}<2.2 \rad$.

Other restrictions are applied to suppress \qq background, which has a two-jet topology 
in contrast to \BB\ events with a more uniform angular distribution of the tracks and 
clusters. 
Events must have $\R2 \leq 0.5$, where \R2 is the second normalized Fox-Wolfram moment
    \cite{Wolfram}, determined from all charged and neutral particles
    in the event.   
We also require $\cosThrust \leq 0.9$, where $\Delta\theta_{\rm thrust}$ is the angle
    between the thrust axis of the $Y$ candidate's decay particles and
    the thrust axis of 
    all other detected particles in the event.  
We require $\L2 < 3.0 \gev$, with   
  $\L2 = \sum_i p^*_i {\cos}^2\theta^*_i$,  
where the sum runs over all tracks in the event excluding the $Y$ candidate, 
and $p^*_i$ and $\theta^*_i$ refer to the c.m.\ momenta and the angles 
measured with respect to the thrust axis of the $Y$ candidate.

We reject candidates
  that have a charged lepton and a low-momentum charged pion consistent with a 
  \btodstarcharges, $D^{*-} \to \Dzb \pi^-_{\rm slow}$ decay 
  as described in \cite{Simonetto2002}.

The kinematic consistency of the candidate decay with a signal $B$
decay is ascertained by restrictions on two variables, the beam-energy
substituted \B~mass \mes, and the difference between the reconstructed
and expected energy of the $B$ candidate \DeltaE. 
In the laboratory frame these variables are defined
as 
$\mES = \sqrt{ (s/2 + \vec{p}_B \cdot \vec{p}_{\epem})\,^2 / E_{\epem}^2 - p_B^2} $
and
$  \DeltaE  = (P_{\epem} \cdot P_{B} - s/2) /  \sqrt{s}$, 
where 
$P_{B}     = (E_{B},        \vec{p}_B)$ and 
$P_{\epem} = (E_{\epem}, \vec{p}_{\epem}) $ are the 
four-momenta of the $B$ meson and the colliding beams, respectively. 
For correctly reconstructed signal
$B$ decays, the \DeltaE distribution is centered at zero, and the \mES distribution peaks at the $B$ mass.  
We restrict candidates to 
$-0.95<\DeltaE<0.95 \gev$ and  $5.095 < \mES <5.295 \gev$.

\subsection{Neural Network Selection}

To separate signal candidates from the remaining background we employ two separate neural networks
(NN), to suppress \qq background and \BtoXclnu background. 
The \qq NN is trained on a sample passing 
the preselection criteria, while the \BtoXclnu NN is
trained on a sample passing both the preselection and the \qq neural network criteria.  
The training is performed with signal and background MC samples.  
These NN are multilayer perceptrons that have two hidden layers with seven and three nodes.

The variables used as inputs to the \qq NN are 
\R2, \L2, \cosThrust, \cosBY, \Mmiss2Emiss, 
Prob$(\chi^2, \Yvtx)$, 
the polar angle of the missing momentum vector in the laboratory frame,
and the Dalitz plot amplitude 
$A_{\rm Dalitz} = \alpha |{\vec p}_{\pi^+} \times {\vec p}_{\pi^-}|$, 
with the $\pi^+$ and $\pi^-$ momenta measured in the \om rest frame and scaled by a
normalization factor $\alpha$.    
True \om mesons typically have larger values of $A_{\rm Dalitz}$ than
combinatorial \om candidates reconstructed from unrelated pions.   
The \BtoXclnu NN uses the same variables, except for \cosThrust, which is replaced by \ctl, 
the helicity angle of the lepton, defined as the angle between the momentum of the lepton 
in the rest frame of the virtual $W$ and the momentum of the $W$ in the rest frame of the $B$.  
The data and MC simulation agree well for the NN input variables at each stage of the
selection.  
The NN discriminators are chosen by
maximizing $\sqrt{\epsilon_{\rm sig}^2 + (1-\eta_{\rm bkg})^2}$, where
$\epsilon_{\rm sig}$ is the efficiency of the signal and 
$\eta_{\rm bkg}$ is the fraction of the background misidentified as signal.  

The selection efficiencies for the various stages of the candidate selection for the signal and background components are given in 
Table \ref{tab:effSummary}. 
After the preselection and NN selection, 
21\% of events in data contribute multiple \Btoomegalnu candidates.
The candidate with the largest value of Prob$(\chi^2, \Yvtx)$ 
is retained.
For the remaining candidates, the reconstructed 3-pion mass 
is required to be consistent with the \om\ nominal mass \cite{PDG2010},
  $|\m3pi - m_\om| < 23 \mev$.  
The overall signal efficiency is 0.73\% if the reconstructed candidate includes a true \om 
and 0.21\% if it includes a combinatorial \om.  
The efficiencies of the \BB and \qq backgrounds are suppressed by several orders of 
magnitude relative to the signal.

\begin{table}[htb]
\begin{center}
\caption
[Efficiency summary]{
  Successive efficiencies (in \%) predicted by MC simulation for each stage of the selection, for
  true- and combinatorial-\om signal, and backgrounds from \BB and \qq events.
}

\begin{tabular}{lrrrr} 
\hline
\hline   
\\[-9pt] 
Source          	&  true-\om  		& comb.-\om  		& \BB 	&  \qq    \\ 
	          	&  signal 			&  signal 	\\  
\hline
Preselection    &    1.9          &  4.8          &  0.0094  &  0.00073  \\
Neural nets      &    43          & 17             & 7.9      & 11      \\
3-pion mass     &    88          & 26             & 24      & 30      \\
\\[-8pt]   
Total (product)          &    0.73 & 0.21  & 0.00018  & 0.000024   \\ 
\hline
\hline
\end{tabular}
\label{tab:effSummary}  
\end{center}
\end{table}

\subsection{Data-MC Comparisons}

The determination of the number of signal events relies heavily on the MC 
simulation to correctly describe the efficiencies and resolutions,
as well as the distributions for signal and 
background sources. Therefore a significant effort has been devoted to 
detailed comparisons of data and MC distributions,
for samples that have been selected to enhance a given source of
background. 

Specifically, we have studied the MC 
simulation of the neutrino reconstruction for a control sample of
\btodstarcharges decays, with 
$\Dstarm \to \Dzb \pi_{\rm slow}^-$ and 
$\Dzb \to K^+ \pi^- \pi^0$.  This final state is similar to that of
the \Btoomegalnu decay, except for the addition of the slow pion
$\pi_s^-$ and the substitution of a $K^+$ for a $\pi^+$.  
This control sample constitutes a high-statistics and high-purity 
sample on which to test the neutrino reconstruction.  
We compare data and MC distributions for the control sample and find good
agreement for the variables used in the preselection 
and as inputs to the NN. We have also used this sample to study
the resolution of the neutrino reconstruction and its impact on \q2,
\mES, and \DeltaE.

\section{Signal Extraction}

\subsection{Fit Method}

We determine the signal yields by 
performing an extended binned maximum-likelihood fit to the observed three-dimensional $\DeltaE$-$\mES$-$q^2$ distributions. 
The fit technique~\cite{Barlow:1993} accounts for the  statistical fluctuations of the data and MC samples.

\begin{figure}
\begin{center}
    \includegraphics[width=0.8\columnwidth]{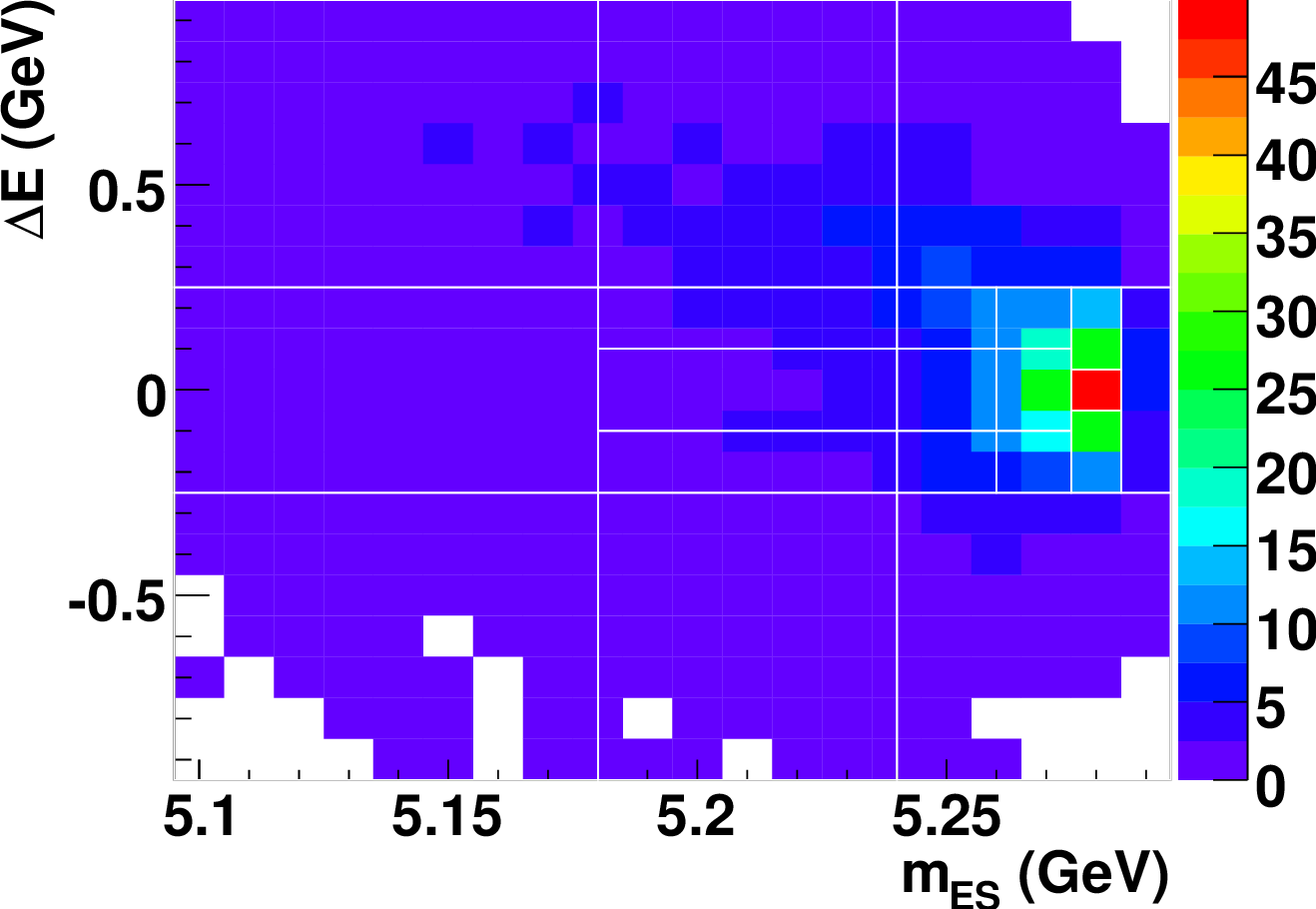}
  \caption{
    Distribution of \DeltaE versus \mES for true-\om signal MC.  
    The 20 bins into which the plane is divided for the fit histogram are overlaid.  
  }
\label{fig:dEmESBins} 
\end{center}
\end{figure}

For this fit the \DeltaE-\mES plane is divided into 20 bins, 
as shown in Fig. \ref{fig:dEmESBins}, and the data are further subdivided 
into five bins in \q2, chosen to contain
roughly equal numbers of signal events. The \q2\ resolution is dominated by the 
neutrino reconstruction. It can be improved
by substituting the missing energy with the magnitude of the missing momentum 
and by rescaling \vpmiss to force $\DeltaE = 0$,
$\q2_{\rm corr} = {[(E_\ell, \vec{p}_\ell) + \delta \cdot (\pmiss, \vpmiss)]}^2$, where 
$ \delta = 1 - \DeltaE / \Emiss$.  
This correction to \q2 is used in the fit.

We describe the measured $\DeltaE$-$\mES$-$q^2$ distribution as a sum of four contributions: 
\Btoomegalnu signal (both true-\om and combinatoric-\om), true-\om \BB, true-\om \qq, and   
the sum of the combinatorial-\om background from \BB and \qq      
events.

\begin{figure}[htbp]
\begin{center}
  \epsfig{file=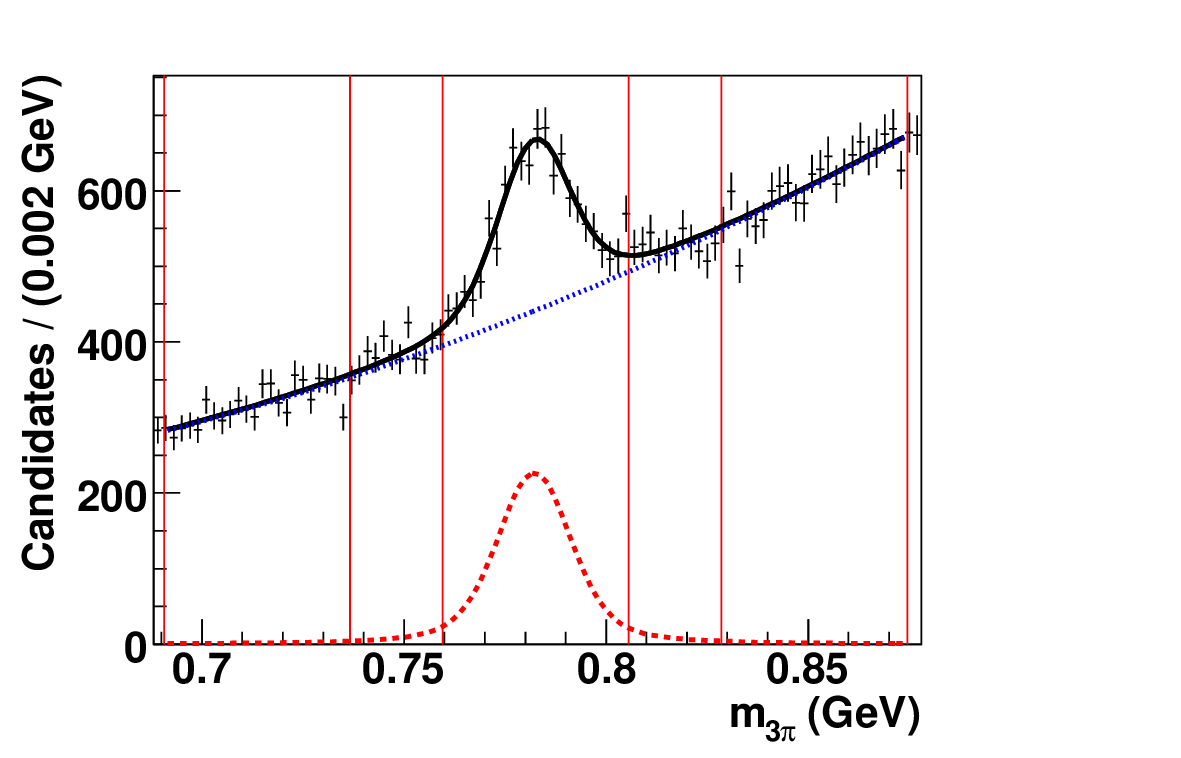, width=\columnwidth}
  \caption{
    Fit to the distribution of \m3pi for data from the all-\q2 sample, 
    with MC combinatorial-\om signal subtracted.  
      The dashed (red) and dotted (blue) curves describe the fitted peaking and
      combinatorial background functions, respectively, and the solid (black) curve is their sum.
      The peak and sideband regions are also indicated.  
      }
\label{fig:m3piFit}
\end{center}
\end{figure}

While the \DeltaE-\mES shapes for the signal and true-\om
\BB and \qq sources are taken from MC samples, we choose to represent
the dominant combinatorial-\om background by the distributions of data
events in the \m3pi sidebands, thereby reducing the dependence on MC
simulation of these backgrounds.  The normalization of these
background data is taken from  
a fit to the 3-$\pi$ mass distribution in the range $0.680 < \m3pi < 0.880$ \gev.
To obtain a sample corresponding to the combinatorial-\om background from \BB and \qq events only,  
we subtract the MC simulated \m3pi contribution of the small combinatorial-\om \Btoomegalnu signal sample.
To the resulting \m3pi distribution we fit 
the sum of a relativistic Breit-Wigner convolved with a normalized Gaussian function, 
and the combinatorial background described by a second degree polynomial.
The resulting fit to the \m3pi distribution for the all-\q2 sample is shown in 
Fig.~\ref{fig:m3piFit}.  
The $\chi^2$ per number of degrees of freedom (dof) for the fits are within the range expected for good fits.
The fitted background function is used to determine 
the weights 
to apply to the upper
and lower sidebands to scale them to the expected yield of combinatorial-\om \BB and \qq background in the \m3pi peak region.

The peak and two sideband regions
are chosen to have a width of 46
\mev and are separated by 23 \mev,
as indicated in Fig.~\ref{fig:m3piFit}.  
Since the normalization of the combinatorial-\om signal contribution depends on the fitted signal yield, 
which is a priori unknown, 
this component is determined iteratively.

The fit has seven free parameters, five for the signal yields in each \q2 bin, 
and one each for the yields of the true-\om \BB and \qq backgrounds, 
the shapes of the distributions are taken from MC simulations.  
The fitted yields are expressed as scale factors relative to the default yields of the MC simulation. 
The total signal yield is taken as the sum of the fitted yields in the 
individual \q2 bins, taking into account correlations.

% results.tex
\subsection{Fit Results}
% See below for how to set up environment.  
% To run fits:  
% wulsin@yakut14~/releases/XSL_ana52/workdir>  scripts/runFit.csh
% 
% To make changes to the fit structure, modify:
% ~/afs/slac/releases/XSL_ana52/EcsConfig/BAD1891/BarlowFit_sdBand-5KVBins_Bp_omegalnu.txt
%
% Use piBFlo as default.  

%\begin{sidewaystable}[htb]
% Get numbers for this table from:  calc/yieldsPostfit5q2VsAll.xlsx.  
%
%
%chi2 value for bin 4 comes from fitTree/fitDataNq2Syst/5q2bins/defaultNew/runStudyFitTree.log  
% logbook 2012-03-31  
\begin{table*}[!htb]
\begin{center}
% Get numbers from: 
% fitTree/fitDataNq2Syst/5q2bins/defaultNew/Barlow.log
% 
%% results/tab_yieldsPostfit5q2vsAll.tex
\caption[Yields from the 5-\q2 bins and all-\q2 fits]{
  Number of events and their statistical uncertainties, as determined from the fit, compared with the number of observed events in data.
%  The fit $\chi^2/$NDF and corresponding probability are also given.  
% The ratio between the sum of the yields in the 5-\q2 bins
%  ($S$) and the yield from the all-\q2 fit ($A$) is given.  The
%  true-\om and combinatorial-\om signal components are fit with  
%  the same scale factor in the fit.  
The combinatorial-\om background yields are fixed in the fit; the quoted uncertainties 
are derived from the sideband subtraction.
%  Components listed in {\bf bold} have scale factors that 
%  % with errors of $\pm$ 0 
%  are fixed in the fit; the uncertainties on their yields are determined separately from the fit.  
}
\begin{tabular}{lrrrrrrr} 
\hline
\hline
\\[-9pt]  
%\q2 bin            			& 1             		& 2            			& 3            			& 4            			& 5            	 		& all \q2 ($A$) 		 \\  
\q2 range (\gevSq) 	     	&  0--4  			&  4--8 			&  8--10 			& 10--12  			& 12--21 			&  0--21       \\ 
\hline
All signal         		 	&  257 $\pm$  72   	&  238 $\pm$ 44 	&  161 $\pm$ 32   	&  177 $\pm$ 32   	&  293 $\pm$ 57   	&  1125 $\pm$ 131  	             \\
\hspace{0.2cm} True-\om signal
&                   		   	   238          		&  209       	&  136          		&  137          		&  168          		&   869             		             \\ 
\hspace{0.2cm} Comb.-\om signal  
&                   			    19         		&   28     		&   25          		&   40           	&  125          		&   256             		             \\
\BBbar (true-\om)  	     	&  105 $\pm$ 19   	&  192 $\pm$ 34 	&  154 $\pm$ 27   	&  195 $\pm$ 34   	&  411 $\pm$ 73   	&  1057 $\pm$ 187  	             \\
\qq    (true-\om)     		&  409 $\pm$ 96   	&  145 $\pm$ 34 &   65 $\pm$ 15     &   34 $\pm$ 8      &   64 $\pm$ 15     &   716 $\pm$ 167  	             \\
Comb.-\om bkgd.  	     	& 1741 $\pm$ 23     & 1818 $\pm$ 24 & 1240 $\pm$ 20     & 1520 $\pm$ 22     & 3913 $\pm$ 35     & 10232 $\pm$ 57      \\
%\hline   
\\[-8pt]   % vertical space
% {\bf Total}       			& 2512          		& 2392         	& 1620         		& 1925         		& 4681         		& 13130           		                \\
Data              			& 2504 $\pm$ 50		& 2433 $\pm$ 49 	& 1605 $\pm$ 40 		& 1858 $\pm$ 43 		& 4738 $\pm$ 69 		& 13138 $\pm$ 115                \\
%\hline
%\\[-9pt]  
%$\chi^2/$NDF   		\\ % 	& 21.7/17       		& 20.6/18     		& 12.7/18    	& 23.6/18       		& 19.4/18       		& 15.2/17    \\
%Prob($\chi^2$, NDF) 	\\ %    &  20\%  			& 30\%   			& 81\%  			& 17\%   			& 37\%  		     	&  58\%   \\  
\hline
\hline
\end{tabular}
\label{tab:yields}
\end{center}
%\end{sidewaystable}
\end{table*}

The fitting procedure has been validated on pseudo-experiments generated
from the MC distributions.  We find no biases and the
uncertainties follow the expected statistical distribution.

\begin{figure*}[!htb] 
  \centering
  \begin{tabular}{ccc}
    \begin{minipage}{0.33\linewidth}
    \centerline{$0<\q2<4 \gevSq$}  \vskip0.1cm
    \epsfig{file=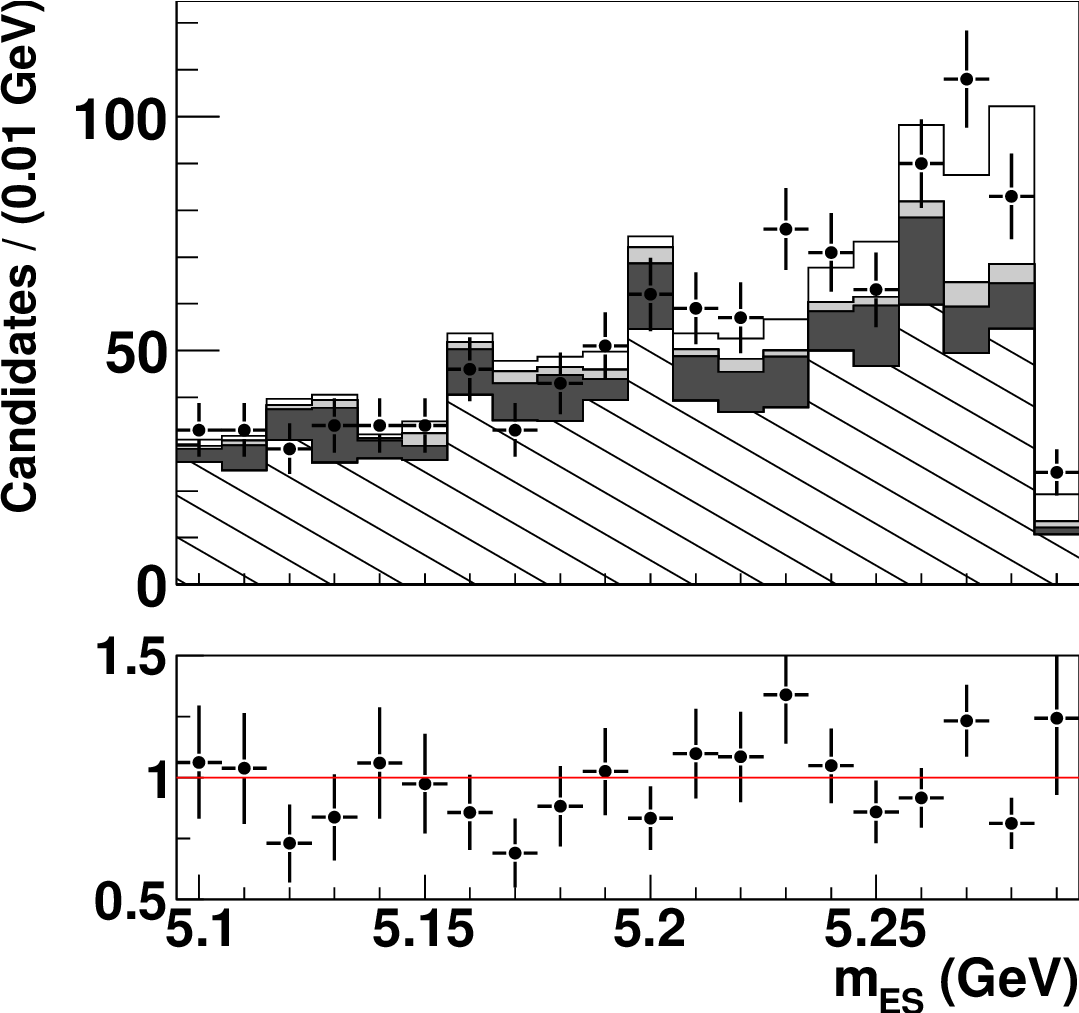, width = 5.2cm}	 \vskip0.6cm
    \centerline{$10<\q2<12 \gevSq$}  \vskip0.1cm
    \epsfig{file=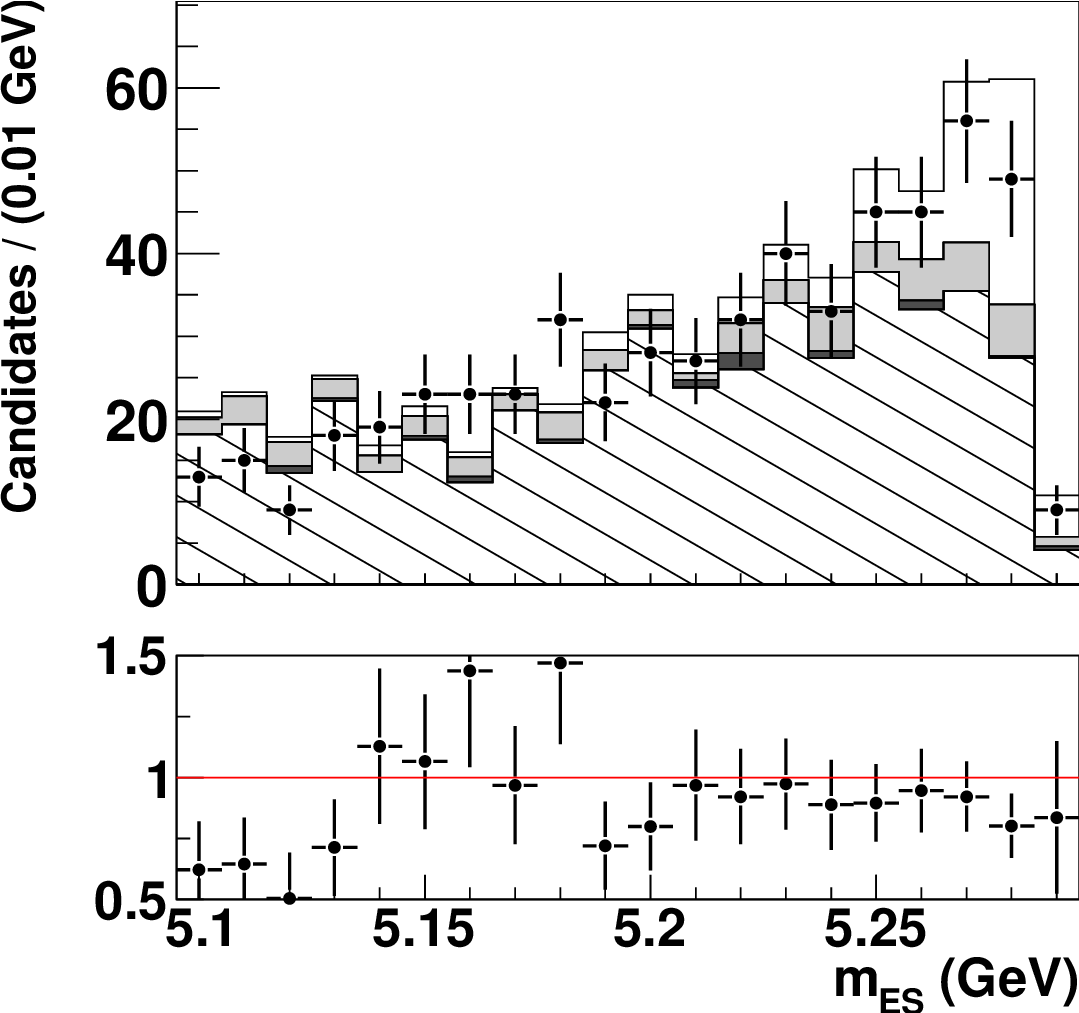, width = 5.2cm} 
    \end{minipage}
    \begin{minipage}{0.33\linewidth}
    \centerline{$4<\q2<8 \gevSq$}  \vskip0.1cm
    \epsfig{file=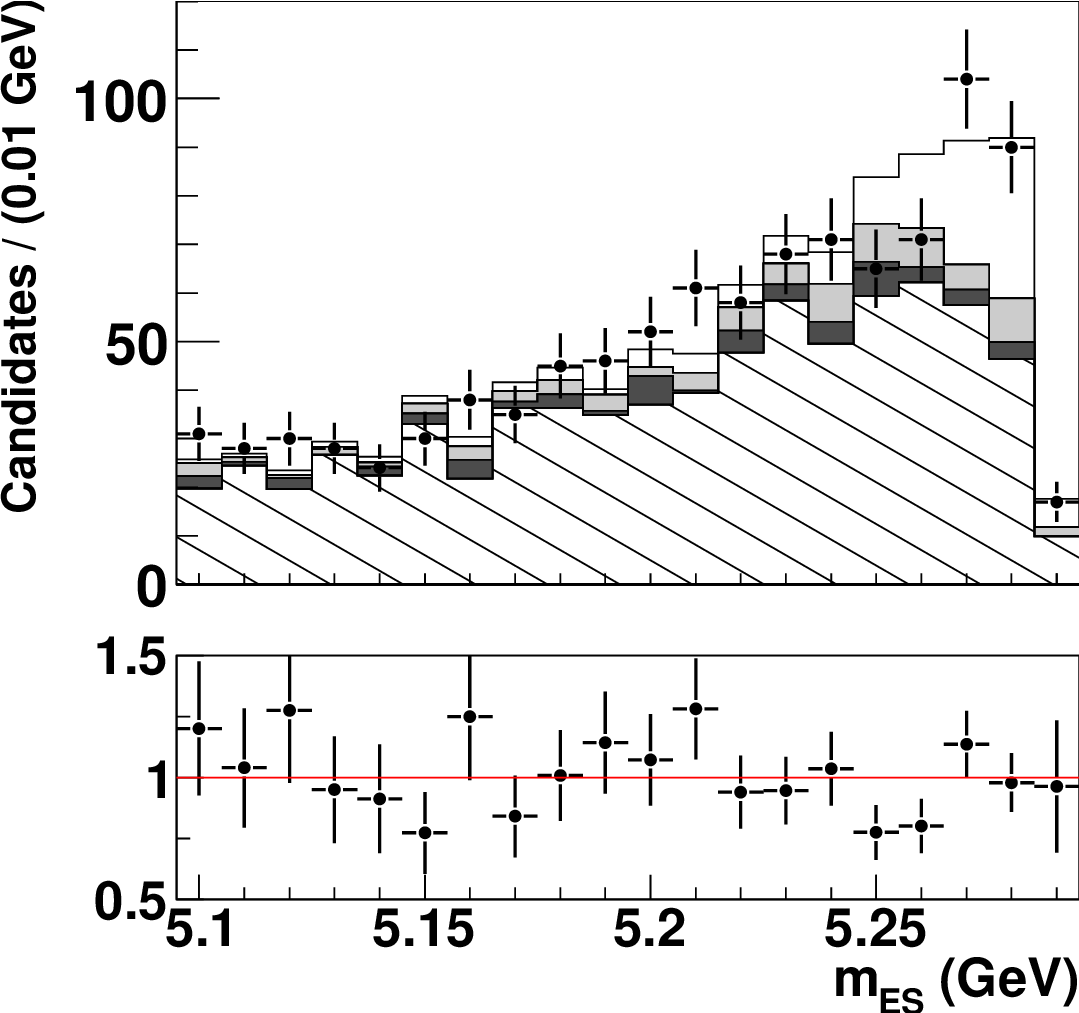, width = 5.2cm}  \vskip0.6cm	
    \centerline{$12<\q2<21 \gevSq$}  \vskip0.1cm
    \epsfig{file=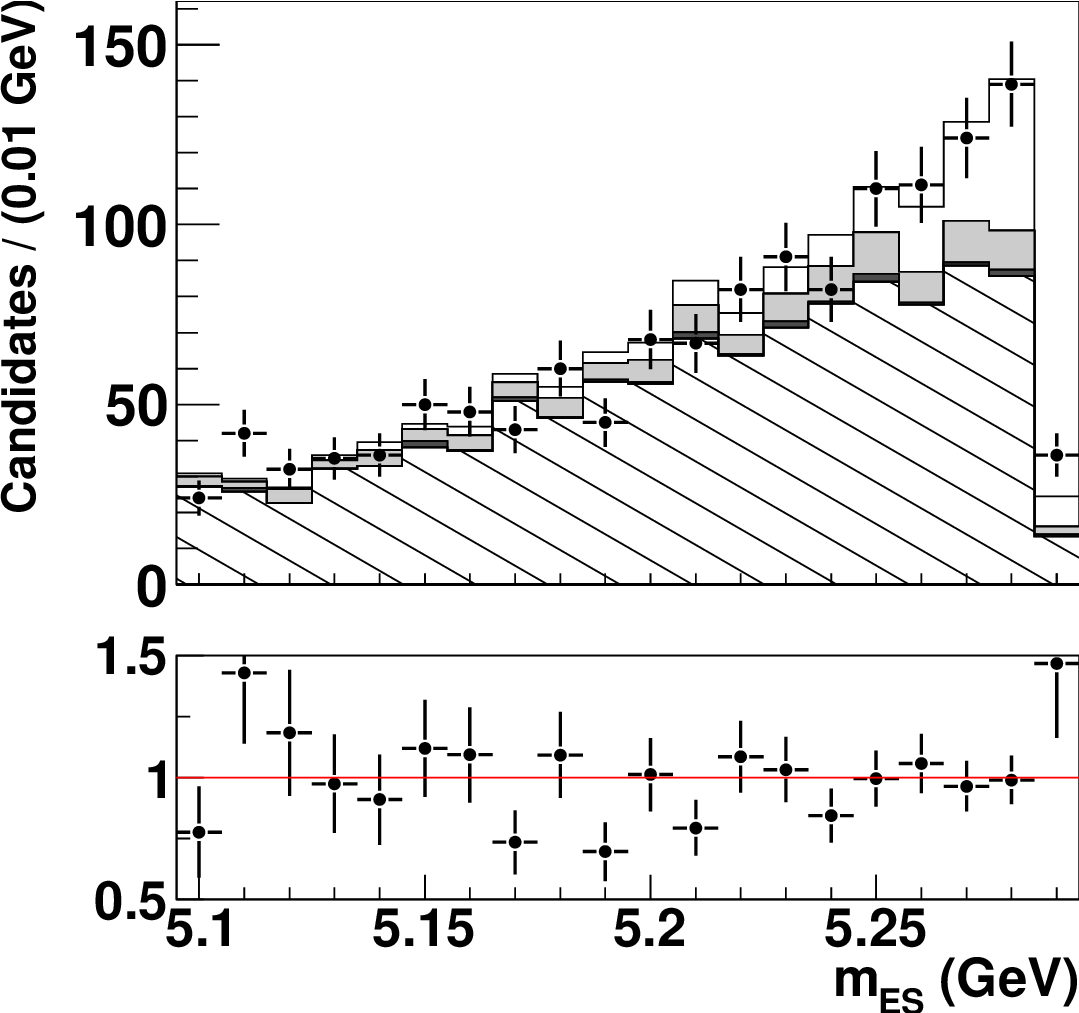, width = 5.2cm}	
    \end{minipage}
    \begin{minipage}{0.33\linewidth}
    \centerline{$8<\q2<10 \gevSq$}  \vskip0.1cm
    \epsfig{file=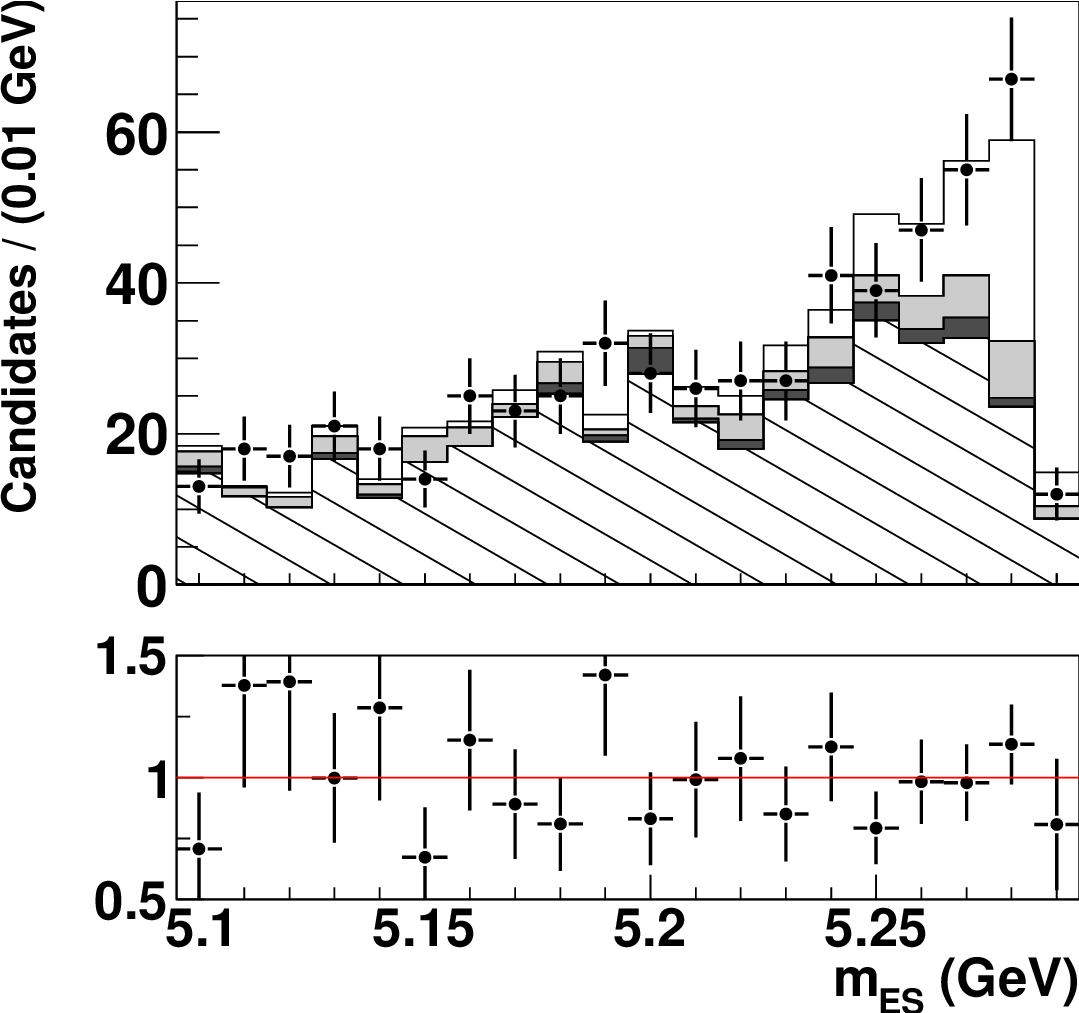, width = 5.2cm}	     \vskip0.6cm
    \centerline{$0<\q2<21 \gevSq$}  \vskip0.1cm
     \epsfig{file=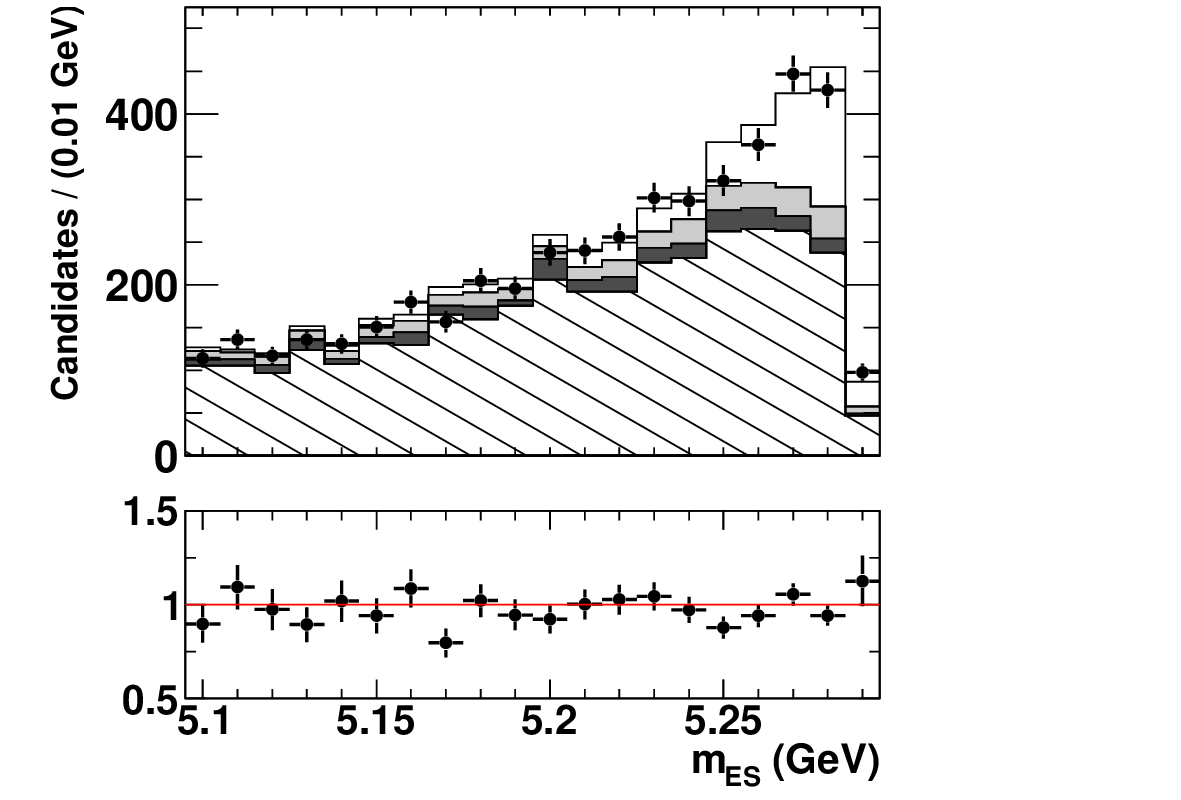, width = 7.5cm}	           
    \end{minipage}
  \end{tabular}
  \caption{
     Distributions of \mES after the fit and the ratio of the data to the fitted predictions, for five separate \q2 bins and the full \q2 range,
     in the \DeltaE signal band, $-0.25< \DeltaE \leq 0.25 \gev$. 
    The points represent data with statistical uncertainties, while the
    stacked histograms
    represent the sum of fitted source components, 
    signal (white), true-\om \BB (light gray), true-\om \qq (dark gray), 
    and combinatorial-\om background (diagonally thatched).  
%    which includes MC
%    contributions and data from the \m3pi sidebands.      
}
  \label{fig:mESFitted}
\end{figure*}

The yields of the signal, true-\om \BB, and true-\om \qq components 
obtained from the binned maximum-likelihood fit to \DeltaE-\mES-\q2 are 
presented in Table \ref{tab:yields}.  
%The fit, performed on a sample with 20 \DeltaE-\mES bins in each of five \q2 bins and seven free parameters, 
%has 93 degrees of freedom.  
%wulsin@noric01~/releases/XSL_ana52/workdir/fitTree> cat fitDataNq2Syst/5q2bins/defaultNew/FitResult.txt
%Chi^2/NDF, no fluc.'s, incl. MC error (default) = 106.204/93 = 1.14198
%root [19] TMath::Prob(106.204, 93)
%(Double_t)1.64989644601007762e-01
%The uncertainties for true-\om \qq in \q2 bins 2--5 are derived as the same 
%fractional uncertainty on \qq in \q2 bin 1.  
%The uncertainties on the combinatorial-\om background are calculated from the \m3pi fit errors.  
%The fitted scale factors are given in Table \ref{tab:fittedParams}; these parameters correspond
%to those listed previously in Table \ref{tab:fitParams3q2}.    
%The signal yield from the fit to the full \q2 range
%is consistent within the uncertainty with the signal yields 
%summed over all five individual \q2 bins.   
%However, the true-\om \qq and \BB yields for the single \q2 fit differ
%significantly from the sum of the five \q2 bins; this can be explained
%by  the large anti-correlation between these two components.    
% (9951) / (12793 - 799) = 83.0\%.    
%It is determined according to the procedure described in Section \ref{sec:FitComponents}, and its yield and distribution are fixed in the \DeltaE-\mES fit.  
%The combinatorial-\om background yield for the all-\q2 fit agrees within 0.1\% with 
%the sum of the yields in the 5 \q2 bins.  
%
Projections of the fitted distributions of \mES for the all-\q2 fit and
for the five \q2 bins fit are shown in Fig.~\ref{fig:mESFitted}.  
The agreement between the data and fitted MC samples
is reasonable for distributions of \DeltaE, \mES, and \q2, as
%, and the unrolled \DeltaE-\mES fit histogram.  
indicated by the 
$\chi^2/{\rm dof}$ of the fit, 106/93, which has a probability of 16\%.  
The fixed combinatorial-\om background yield accounts for 83\% of all backgrounds. 
%
%$\chi^2/$NDF probabilities listed in Table \ref{tab:yields}.
%The $\chi^2$ probabilities are all reasonable, ranging from 18.0\% in bin 4 to 81\% in bin 3; for the all-\q2 fit the probability is 58\%.   
%The number of degrees of freedom are given by 
%NDF $= N_{\rm bins} - N_{\rm par}$, 
%where 
%$N_{\rm bins}$ is the number of \DeltaE-\mES fit bins, and 
%$N_{\rm par}$ is the number of free parameters in the fit (3 for all-\q2 and \q2 bin 1; 2 for \q2 bins 2-5).  
%Bins in the fitted histograms with fewer than 10 entries for either data or MC are retained in the fit, but 
%are not included in the number of degrees of freedom or in the calculation of $\chi^2$.   
%  (this only affects bin 4).  
%In each bin, the $\chi^2$ is lower after the fit, as expected.  
%Also listed is the $\chi^2$ probability, 
%defined as the probability that an observed $\chi^2$ would exceed
%the measured $\chi^2$ by chance.  
%
%
The correlations among the parameters are listed in Table
\ref{tab:fitCorr}.  The strongest correlation is $-72\%$, between the signal and \qq yields 
in the first \q2 bin, which contains most of the \qq background.  
The correlation between signal and \BB background is strongest in the last \q2 bin, $-40\%$, 
because of a large contribution from other \BtoXulnu decays.  
Correlations among signal yields are significantly smaller.  

% Table from results.tex
\begin{table}[!htb]
% Values taken from:
% /releases/XSL_ana50/workdir/fitTree> e fitDataNq2Syst/5q2bins/defaultNew/Barlow.log
% PARAMETER  CORRELATION COEFFICIENTS
%       NO.  GLOBAL      6      7      8      9     10     11     12
%        6  0.80278   1.000 -0.466 -0.724 -0.106 -0.031  0.051  0.088
%        7  0.70682  -0.466  1.000  0.223 -0.249 -0.253 -0.284 -0.401
%        8  0.73595  -0.724  0.223  1.000  0.121  0.061  0.001 -0.011
%        9  0.35386  -0.106 -0.249  0.121  1.000  0.105  0.094  0.128
%       10  0.30409  -0.031 -0.253  0.061  0.105  1.000  0.088  0.121
%       11  0.29902   0.051 -0.284  0.001  0.094  0.088  1.000  0.125
%       12  0.41679   0.088 -0.401 -0.011  0.128  0.121  0.125  1.000

\centering
\caption[Correlations among the fit parameters]{
  Correlations among the fit scale factors $p^s_k$ for the simulated source 
  $s$ and \q2 bin $k$.  The scale factors for \qq and \BB apply to the full 
  \q2 range.  
%  All empty entries are zero; there are no  
%  correlations between separate \q2 bins. 
}
\begin{tabular}{lrrrrrrrr} 
\hline
\hline
\\[-9pt]
& 
& $p^{\qq}$ 
& $p^{\BB}$ 
& $p_1^{\om\ell\nu}$ 
& $p_2^{\om\ell\nu}$ 
& $p_3^{\om\ell\nu}$  
& $p_4^{\om\ell\nu}$  
& $p_5^{\om\ell\nu}$  \\ 
\hline
\\[-9pt]  
%& $p^{\qq}$          &    1.000 & $-$0.466 & $-$0.724 & $-$0.106 & $-$0.031 &    0.051 &    0.088 \\
%& $p^{\BB}$          & $-$0.466 &    1.000 &    0.223 & $-$0.249 & $-$0.253 & $-$0.284 & $-$0.401 \\
%& $p_1^{\om\ell\nu}$ & $-$0.724 &    0.223 &    1.000 &    0.121 &    0.061 &    0.001 & $-$0.011 \\
%& $p_2^{\om\ell\nu}$ & $-$0.106 & $-$0.249 &    0.121 &    1.000 &    0.105 &    0.094 &    0.128 \\
%& $p_3^{\om\ell\nu}$ & $-$0.031 & $-$0.253 &    0.061 &    0.105 &    1.000 &    0.088 &    0.121 \\
%& $p_4^{\om\ell\nu}$ &    0.051 & $-$0.284 &    0.001 &    0.094 &    0.088 &    1.000 &    0.125 \\
%& $p_5^{\om\ell\nu}$ &    0.088 & $-$0.401 & $-$0.011 &    0.128 &    0.121 &    0.125 &    1.000 \\ 

& $p^{\qq}$         &    1.000 & $-$0.466 & $-$0.724 & $-$0.106 & $-$0.031 &    0.051 &    0.088 \\
& $p^{\BB}$         &          &    1.000 &    0.223 & $-$0.249 & $-$0.253 & $-$0.284 & $-$0.401 \\
& $p_1^{\om\ell\nu}$ &          &          &    1.000 &    0.121 &    0.061 &    0.001 & $-$0.011 \\
& $p_2^{\om\ell\nu}$ &          &          &          &    1.000 &    0.105 &    0.094 &    0.128 \\
& $p_3^{\om\ell\nu}$ &          &          &          &          &    1.000 &    0.088 &    0.121 \\
& $p_4^{\om\ell\nu}$ &          &          &          &          &          &    1.000 &    0.125 \\
& $p_5^{\om\ell\nu}$ &          &          &          &          &          &          &    1.000 \\

% With minus signs in math-mode the table is too wide.  
% So I only use two digits after the decimal.  
%$p^{\qq}$     & $-0.739$ & $-0.488$ &        &        &        &        &        &        \\  
%$p^{\BB}$     &  1.000 &  0.016 &        &        &        &        &        &        \\
%$p^{\qq}_1$  &        &         & $-0.733$ & $-0.511$ &        &        &        &        \\ 
%$p^{\BB}_1$  &        &        &  1.000 & $-0.018$  &        &        &        &        \\
%$p^{\BB}_2$  &        &        &        &        & $-0.673$ &        &        &        \\
%$p^{\BB}_3$  &        &        &        &        &        & $-0.651$ &        &        \\
%$p^{\BB}_4$  &        &        &        &        &        &        & $-0.593$ &        \\
%$p^{\BB}_5$  &        &        &        &        &        &        &        & $-0.577$ \\
% $p^{\qq}$     & $-0.74$ & $-0.49$ &        &        &        &        &        &        \\  
% $p^{\BB}$     &  1.00 &  0.02 &        &        &        &        &        &        \\
% $p^{\qq}_1$  &        &         & $-0.73$ & $-0.51$ &        &        &        &        \\ 
% $p^{\BB}_1$  &        &        &  1.00 & $-0.02$  &        &        &        &        \\
% $p^{\BB}_2$  &        &        &        &        & $-0.67$ &        &        &        \\
% $p^{\BB}_3$  &        &        &        &        &        & $-0.65$ &        &        \\
% $p^{\BB}_4$  &        &        &        &        &        &        & $-0.59$ &        \\
% $p^{\BB}_5$  &        &        &        &        &        &        &        & $-0.58$ \\
\hline
\hline
\end{tabular}
\label{tab:fitCorr}  
\end{table}

The branching fraction, \BR(\Btoomegalnu), averaged over electron and muon channels,
is defined as 
$  \BR(\Btoomegalnu)
  = \sum_i (N_i^{\mathrm{sig}} / \epsilon_i^{\mathrm{sig}})/(4  f_{\pm} N_{\BB})$, 
%\begin{equation}
%  \BR(\Btoomegalnu)
%  = {N_{\Btoomegalnu} \over 4 \times N_{\upsbpbm}} 
%  = {N_{\rm sig}^{\rm reco} \over 4 \times \epsilon_{sig} f_{\pm} N_{\BB}},
%\end{equation}
where $N_i^{\rm sig}$ refers to the number of reconstructed electron and muon signal events
in \q2 bin $i$, 
$\epsilon_i^{\rm sig}$ is the reconstruction efficiency, 
%which takes into account the effect of finite \q2 resolution, 
$f_{\pm}$
% = 1-f_{00} = 0.516 \pm 0.006$
% = 1 - 0.484 \pm 0.006$ 
% \cite{PDG2010}  
is the fraction of \BpBm decays in all \BB events, and $N_{\BB}$
is the number of produced \BB events.
% produced in the \epem interaction.  
The factor of 4 comes from the
fact that ${\cal B}$ is quoted as the average of $\ell = e$ and $\mu$ samples, not the sum, 
and the fact that either of the 
two $B$ mesons in the \BpBm event may decay into the signal mode.  
The \q2 resolution in the signal region is 0.36 \gevSq, smaller than the width of the \q2 bins.  
To account for the finite \q2 resolution, the background-subtracted, efficiency-corrected spectrum 
is adjusted by deriving from the signal MC the ratio of the true and reconstructed \q2 spectra, 
$({\rm d} {\cal B}/{\rm d}\q2_{\rm true}) / ({\rm d} {\cal B}/{\rm d}\q2_{\rm reco})$. 
This ratio differs by 9\% at low \q2 and considerably less at higher \q2.  
The partial and total branching fractions listed in Table \ref{tab:BF5q2}
are corrected for the effects of finite \q2 resolution and efficiency.

%Several stability checks are performed to test the sensitivity of the
%results to changes in the fit configuration and the event sample.  

% Values calculated in:
% calc/systSummaryVera.xlsx  
% Any change to this table must also be reflected in a change to:
% * Figure \ref{fig:BFBallVsIsgw2}  
% * Table VI:  Vub
% * chi2 agreement with LCSR & ISGW2 in text  
% abstract  
% conclusion  
\begin{table}[htb]
\centering
\caption
[Measured \Btoomegalnu branching fraction]
{Measured \Btoomegalnu branching fraction and partial
  branching fractions in bins of \q2 with statistical and systematic
  uncertainties.  
}    
\begin{tabular}{cc} 
\hline
\hline
\\[-9pt]  
\q2 (\gevSq)  &  $\Delta{\cal B}$ $(\times 10^{-4}$) \\  
\hline
% deltaB(i) = fitFactor(i) * frac(i) * BFtot_MC   /////  total error
0--4        & 0.214 $\pm$ 0.060  $\pm$ 0.024 \\        % 0.065
4--8        & 0.200 $\pm$ 0.037  $\pm$ 0.010 \\        % 0.038
8--10       & 0.147 $\pm$ 0.029  $\pm$ 0.010 \\        % 0.031
10--12      & 0.169 $\pm$ 0.031  $\pm$ 0.010 \\        % 0.032
12--21      & 0.482 $\pm$ 0.093  $\pm$ 0.038 \\        % 0.101
%\hline			 	    		
\\[-8pt]   % vertical space
0--12      	& 0.730 $\pm$ 0.083  $\pm$ 0.054 \\        % 0.099  Copied from calc/vub.ods.  
0--21      	& 1.212 $\pm$ 0.140  $\pm$ 0.084 \\        % 0.163 
% According to the BaBar style guide, errors should be quoted to 2
% significant digits if the error is less than 0.45, but 1 sig
% digit otherwise.  This suggests that I leave the all-q2 errors as
% 1.5 and 1.2, and then the other errors should be given to the same
% level of accuracy.  
\hline
\hline
\end{tabular}
\label{tab:BF5q2}  
\end{table}

\section{Systematic Uncertainties}

Table \ref{tab:systSumm} summarizes the contributions to the systematic uncertainty.  
The event reconstruction systematic uncertainties are most sensitive to the neutrino reconstruction, which depends on the
detection of all of the particles in the event. To assess the impact of the uncertainty of the measured efficiencies for charged tracks, the MC signal and background samples are reprocessed and the analysis is repeated, after tracks have been eliminated at random with a probability determined by the uncertainty in the tracking efficiency.
Similarly, we evaluate the impact from uncertainties in the photon
reconstruction efficiency by
eliminating photons at random as a function of the photon energy.
Since a \KL leaves no
track and deposits only a small fraction of its energy in the calorimeter,
the reconstruction of the neutrino is impacted.
The uncertainty on the  \KL MC simulation involves
the shower energy deposited by the \KL in the calorimeter,
the \KL detection efficiency, and
the inclusive \KL production rate as a function of momentum from \BB events.

The impact of the changes to the simulated background distributions  which enter the fit
are smaller than for the signal, since the large combinatorial backgrounds
are taken from data, rather than MC simulations.
As an estimate of the impact of these variations of the MC simulated distributions
on the $q^2$ dependent signal yield, we combine the observed reduction in the
signal distribution with the impact of the changes to \qq and \BB backgrounds
on the signal yield, taking into account the correlations obtained from
the fit (see Table III).  Since the correlations between signal and backgrounds
are small at high  $q^2$, the impact of the uncertainties in the background
are also modest. This procedure avoids large statistical
fluctuations of the fit procedure that have been observed to be larger than
the changes in the detection efficiencies.
However, this procedure does not account for the small changes in the shape of the
distributions, and we therefore sum the magnitude of the changes for signal and background,
rather than adding them in quadrature or taking into account the signs of the
correlations of the signal and backgrounds in a given $q^2$ bin.  

We assign an uncertainty on the identification efficiency of electrons and muons, 
as well as on the lepton and kaon vetoes of the \om daughter pions,  
based on the change in signal yield after varying the selector efficiencies 
within their uncertainties.

The uncertainty in the calculation of the LCSR form factors impacts the uncertainty 
on the branching fraction 
because it affects the predicted \q2 distribution of the signal and thereby the fitted signal yield.  
We assess the impact by varying the form factors within their uncertainties.  
We include the uncertainty on the branching fraction of the \om decay, 
\BR(\om $\rightarrow \pi^+\pi^-\pi^0$) = $( 89.2 \pm 0.7 ) \times
 10^{-2}$ \cite{PDG2010}.  
 %, an uncertainty of 0.8\%.  
To evaluate the uncertainty from radiative corrections, candidates are
reweighted by 20\% of the difference
between the spectra with and without PHOTOS~\cite{photos}, which models the final state
radiation of the decay.  

The uncertainty on the true-\om backgrounds has a small impact on the signal yield 
since these components represent
a small fraction of the total sample.   
To assess the uncertainty of the \DeltaE-\mES-\q2 shapes of the true-\om
\qq and true-\om \BB samples, the fit is repeated after the events are
reweighted to reproduce the inclusive \om momentum distribution measured 
in \BB and \qq events.
We also assess the uncertainty on the modeling of the semileptonic backgrounds by 
varying the branching fractions and form factors of 
the exclusive and inclusive \BtoXulnu~\cite{PDG2010} 
and \BtoXclnu backgrounds~\cite{HFAG2010} within their uncertainties.

To assess the uncertainties that result from 
the MC prediction of the \m3pi distribution of the combinatorial-\om signal, 
we use the uncorrected distribution, in which the combinatorial-\om signal is not
subtracted from the \m3pi 
sidebands, and the signal fit parameter is set to scale only the true-\om
signal contribution.  
Twenty percent of the difference between the nominal and uncorrected
results is taken as the systematic
uncertainty;   
it is largest for $12<\q2<21 \gevSq$ because the fraction of combinatorial-\om signal in
this \q2 bin is large.
The sideband event yields determined from the \m3pi fit 
are varied within their fit errors to determine the statistical uncertainty
on the combinatorial-\om background.  
The uncertainty in the chosen \m3pi ansatz is assessed by 
repeating the \m3pi fits, replacing the nominal functions for the peak and background
components. 
For the background component, we use a third  
instead of a second degree polynomial.  
For the peaking component, we use a Gaussian function in place of a
relativistic Breit-Wigner convoluted with a Gaussian function.  
The systematic error from the \m3pi ansatz is taken as the sum in
quadrature of the change in signal yield for each of these functional
variations.

The branching fraction depends inversely on the value of $N_{\BB}$, 
which is determined with a
precision of $1.1\%$ \cite{bcount}.
At the \FourS\ resonance, the fraction of  \BpBm events
is measured to be 
$f_{\pm} = 0.516 \pm 0.006$~\cite{PDG2010}, 
with an uncertainty of 1.2\%.

\begin{table}[htb]
\centering
\caption{Systematic uncertainties in \% on the branching fraction.}  
\begin{tabular}{lrrrrrr} 
\hline
\hline
\\[-9pt]  
\q2 range (\gevSq)                                             &  0--4 &  4--8 &  8--10& 10--12& 12--21& 0--21 \\ 
\hline
\\[-8pt]   
{\bf Event reconstruction}         \\
\hspace{0.0cm} Tracking efficiency                             &  3.9  &  1.5  &  2.8  &  2.3  &  1.1  &  2.0  \\
\hspace{0.0cm} Photon efficiency                               &  2.0  &  1.7  &  3.3  &  1.1  &  0.6  &  1.5  \\ 
\hspace{0.0cm} $K_L$ prod./interactions                        &  4.8  &  1.8  &  2.5  &  1.1  &  1.4  &  1.9  \\
\hspace{0.0cm} Lepton identification                           &  1.6  &  1.5  &  1.5  &  1.2  &  1.2  &  1.3  \\
\hspace{0.0cm} $K/\ell$ veto of \om daughters                  &  1.7  &  1.7  &  1.7  &  1.7  &  1.8  &  1.7  \\
\\[-8pt]   
{\bf Signal simulation}         \\
\hspace{0.0cm} Signal form factors                             &  6.3  &  1.5  &  1.1  &  2.9  &  4.6  &  4.8 \\
\hspace{0.0cm} \BR(\om $\rightarrow \pi^+\pi^-\pi^0$)          &  0.8  &  0.8  &  0.8  &  0.8  &  0.8  &  0.8 \\
\hspace{0.0cm} Radiative corrections                           &  0.4  &  0.3  &  0.2  &  0.1  &  0.2  &  0.2 \\                                                                   
\\[-8pt]   
{\bf True-\om background}         \\                                                                                  
\hspace{0.0cm} \qq \DeltaE-\mES-\q2 shapes                     &  2.6  &  0.1  &  0.4  &  0.2  &  0.3  &  0.5 \\
\hspace{0.0cm} \BB \DeltaE-\mES-\q2 shapes                     &  2.0  &  0.9  &  1.8  &  0.2  &  0.1  &  0.8 \\
\hspace{0.0cm} \BtoXclnu\ \BR\ and FF                          &  0.2  &  0.6  &  0.3  &  0.2  &  0.2  &  0.2 \\
\hspace{0.0cm} \BtoXulnu\ \BR\ and FF                          &  0.3  &  0.4  &  0.4  &  0.3  &  0.5  &  0.4 \\
\\[-8pt]   
{\bf Comb.-\om sources} \\                                                                                  
\hspace{0.0cm}  Signal \m3pi distribution                      &  0.6  &  0.5  &  0.4  &  1.1  &  3.7  &  1.5 \\
\hspace{0.0cm} Bkgd. yield, stat. error                        &  4.2  &  1.0  &  0.9  &  0.9  &  2.0  &  1.7 \\
\hspace{0.0cm} Bkgd. yield, ansatz error                       &  1.7  &  2.2  &  2.7  &  2.7  &  3.5  &  0.9 \\
\\[-8pt]   
{\bf $B$ production}         \\     
\hspace{0.0cm} \BBbar counting                                 &  1.1  &  1.1  &  1.1  &  1.1  &  1.1  &  1.1 \\
\hspace{0.0cm} $f_\pm$                                         &  1.2  &  1.2  &  1.2  &  1.2  &  1.2  &  1.2 \\
\\[-8pt]   
{\bf Syst. uncertainty}                                        & 11.1  &  5.2  &  6.8  &  5.8  &  7.9  &  6.9 \\     
{\bf Stat. uncertainty}      	                               & 28.1  & 18.7  & 20.0  & 18.1  & 19.4  & 11.6 \\ 
{\bf Total uncertainty}                                        & 30.2  & 19.4  & 21.1  & 19.0  & 20.9  & 13.5 \\
\hline
\hline
\end{tabular}
\label{tab:systSumm}  
\end{table}

 \section{Results and conclusions}
\begin{figure}[h]
  \begin{center}
    \begin{minipage}{\columnwidth}
      \includegraphics[width=\columnwidth]{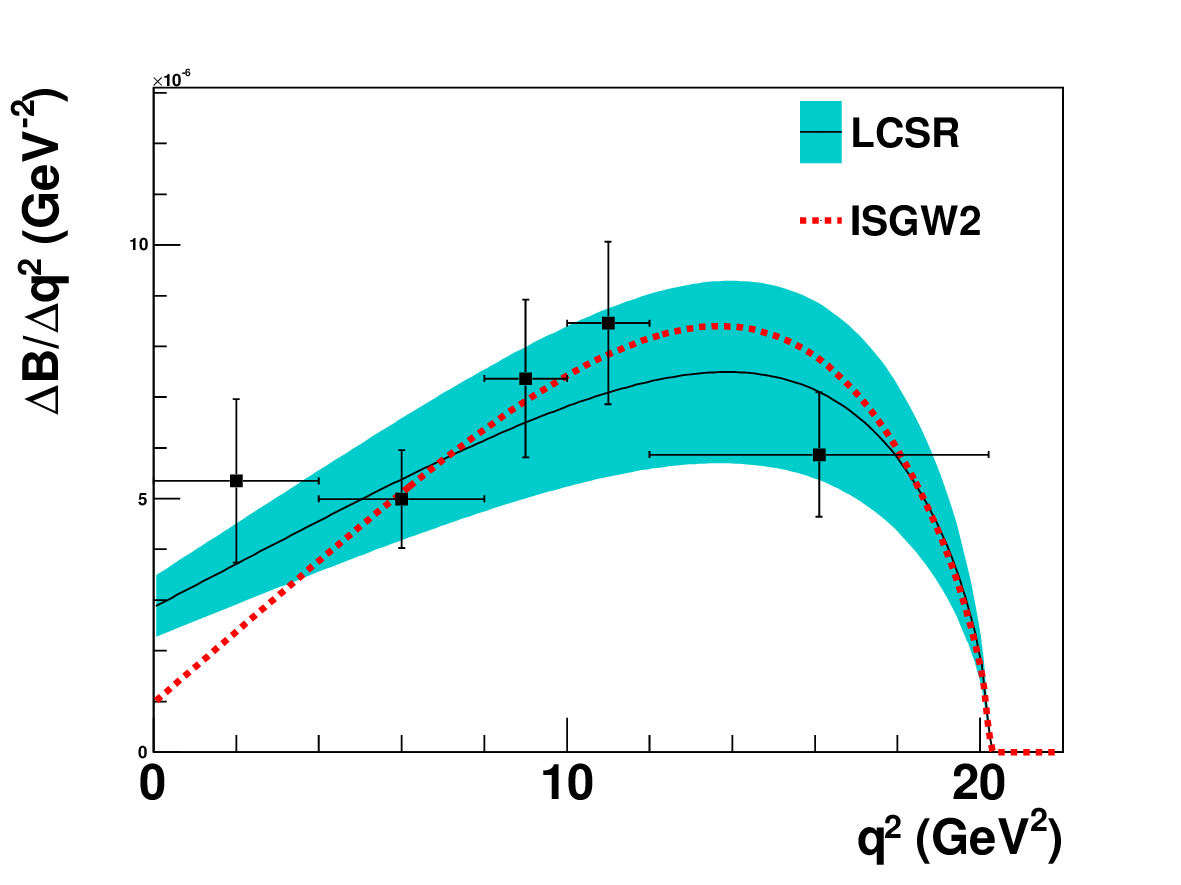}
    \end{minipage}
    \caption[Measured partial branching fractions]{
      Partial branching fractions (points with error bars) with respect to \q2.
      The data are compared with
      the predictions from light-cone sum rules (LCSR) \cite{Ball05} and
      a quark-model calculation (ISGW2) \cite{isgw2}. 
	The uncertainty band (shaded) is given for the LCSR calculation.
    }
    \label{fig:BFBallVsIsgw2} 
  \end{center}
\end{figure}

We have measured the branching fraction,
\begin{eqnarray}
\BR(\Btoomegalnu) = (1.21 \pm 0.14 \pm 0.08) \times 10^{-4}, 
\end{eqnarray}
where the first error is statistical and the second is systematic, 
based on 1125 $\pm$ 131 observed signal candidates.  
Here, $\ell$ indicates the electron or muon
decay mode and not the sum over them.  
The measured partial branching fractions 
are presented in Table \ref{tab:BF5q2} and  
are compared to
the predictions from two form factor calculations in Fig. 
\ref{fig:BFBallVsIsgw2}.  These QCD predictions have been normalized to 
the measured branching fraction.

Neglecting the theoretical uncertainties, the $\chi^2$/NDF of the measured distribution 
relative to the LCSR prediction \cite{Ball05} is 2.4/4, corresponding to a $\chi^2$ probability of 67\%; 
relative to the ISGW2 prediction \cite{isgw2} the $\chi^2$/NDF is 4.2/4, with a $\chi^2$ probability of 40\%.  
Within the large experimental uncertainties,
both the LCSR and ISGW2 form factor 
calculations are consistent with the data.  
The uncertainties of the ISGW2 form factor calculation are not available.  
The uncertainties of the LCSR calculation were estimated 
by the authors
to vary linearly as a function of \q2; i.e., 
$\sigma_{{\rm d} \BR/{\rm d}\q2} / ({\rm d} \BR/{\rm d}\q2) = 21\% + 3\% \times \q2 / (14 \gevSq)$, 
for the \Btorholnu decays~\cite{ZwickyPrivate}.
It is assumed that this estimate is also valid for \Btoomegalnu\ decays.  

The value of \Vub can be determined from the measured partial branching
fraction, the 
$B^+$~lifetime $\tau_+ = (1.638 \pm 0.011) \ps$ \cite{PDG2010}, 
and the integral $\Delta\zeta$
of the predicted differential decay rate:  
  
\begin{eqnarray}
  \label{eq:Vub}
  \Vub &=& \sqrt{\frac{\Delta{\cal B}(q^2_{\rm min},q^2_{\rm max})}
    {\tau_+ \ \Delta\zeta(q^2_{\rm min},q^2_{\rm max})}}   \ , \nonumber  \\ 
  \Delta\zeta(\q2_{\rm min},\q2_{\rm max}) &=& {1 \over \Vub^2} 
  \int_{\q2_{\rm min}}^{\q2_{\rm max}}{ {\rm d} \Gamma_{\rm theory} \over  {\rm d} \q2}
      {\rm d} \q2 \ .  
\end{eqnarray}

Table \ref{tab:vub} lists the values of
$\Delta\zeta$ and \Vub for 
LCSR and ISGW2 in different ranges of \q2.  
LCSR calculations are more accurate at low \q2, while ISGW2 predictions are 
more reliable at high \q2. 
Both form factor calculations arrive at very similar values for \Vub. 
These values of \Vub are consistent with 
the more precisely measured values from \Btopilnu decays \cite{Luth:AnnRev}.  

\begin{table}[htb]
  \centering
  \caption[Extracted value of \Vub]{
	\Vub, determined from two form factor calculations of $\Delta\zeta$, in different
	ranges of \q2.  The first uncertainty is experimental (the sum in quadrature of
	statistical and systematic); the second uncertainty is from theory, and is only 
	available for LCSR.  
      }  
\begin{tabular}{lccll} 
\hline
\hline
\\[-9pt]  
  & \q2 (\gevSq)  &  $\Delta\zeta (\ps^{-1})$ & \Vub ($\times 10^{-3}$) \\  
\hline
& 0--12   &   3.9 $\pm$ 0.9 & 3.37 $\pm$ 0.23 $\pm$ 0.38 \\  
LCSR  \cite{Ball05} 
& 12--21  &   3.2 $\pm$ 0.8 & 3.04 $\pm$ 0.32 $\pm$ 0.37 \\  
& 0--21   &   7.1 $\pm$ 1.7 & 3.23 $\pm$ 0.22 $\pm$ 0.38 \\  
\\[-8pt]   
& 0--12   &   3.6 	& 3.51 $\pm$ 0.24 \\  
ISGW2 \cite{isgw2}  
& 12--21  &   3.4 	& 2.94 $\pm$ 0.31 \\  
& 0--21   &   7.0 	& 3.24 $\pm$ 0.22 \\  
\hline
\hline
\end{tabular}
      \label{tab:vub}  
\end{table}

The value of \BR(\Btoomegalnu) measured in this analysis supersedes the 
previous \babar\ measurement \cite{Anders} based on a smaller data sample, 
and is in excellent agreement with a recent result~\cite{Lees:2012vv} based on the full \babar\ data set.  
The principal difference between this analysis and the previous ones 
is that the combinatorial-\om background 
is taken from the sideband of the data \m3pi distribution rather
than from MC simulation.  
Although the dominant systematic uncertainties from event
reconstruction cannot be avoided, this procedure substantially reduces 
the reliance on the MC simulation
of this largest source of background.

Currently, the QCD predictions of the form factors, and in particular their uncertainties, 
have limited precision for
\Btoomegalnu and \Btorholnu decays.  
These form factor uncertainties impact 
\Vub derived from \BR(\Btoomegalnu).  
In the future, form factor calculations with reduced uncertainties  
combined with improved branching fraction measurements would enable tests and 
discrimination among different predictions as a function of \q2, 
and thereby improve the determination of \Vub.

\section{Acknowledgments}

We are grateful for the 
extraordinary contributions of our \pep2\ colleagues in
achieving the excellent luminosity and machine conditions
that have made this work possible.
The success of this project also relies critically on the 
expertise and dedication of the computing organizations that 
support \babar.
The collaborating institutions wish to thank 
SLAC for its support and the kind hospitality extended to them. 
This work is supported by the
US Department of Energy
and National Science Foundation, the
Natural Sciences and Engineering Research Council (Canada),
the Commissariat \`a l'Energie Atomique and
Institut National de Physique Nucl\'eaire et de Physique des Particules
(France), the
Bundesministerium f\"ur Bildung und Forschung and
Deutsche Forschungsgemeinschaft
(Germany), the
Istituto Nazionale di Fisica Nucleare (Italy),
the Foundation for Fundamental Research on Matter (The Netherlands),
the Research Council of Norway, the
Ministry of Education and Science of the Russian Federation, 
Ministerio de Ciencia e Innovaci\'on (Spain), and the
Science and Technology Facilities Council (United Kingdom).
Individuals have received support from 
the Marie-Curie IEF program (European Union), the A. P. Sloan Foundation (USA) 
and the Binational Science Foundation (USA-Israel).

\bibliographystyle{apsrev}

\bibliography{paper} 

\begin{thebibliography}{30}
\bibitem[{chg()}]{chgConjugate}
\bibinfo{note}{The charge conjugate decay mode is implicitly included.}
\end{thebibliography}

\clearpage

\end{document}